\documentclass[iop]{emulateapj}
\usepackage{amsmath,graphics,epsfig}
\usepackage{natbib,hyperref}
\usepackage{xcolor,soul}
\defcitealias{somers14}{SP14}
\defcitealias{somers15}{SP15}

\begin{document}

\title{Older and Colder: The impact of starspots on pre-main sequence stellar evolution}

\shorttitle{Starspots on the pre-main sequence}

\author{Garrett Somers \& Marc H. Pinsonneault}

\affil{Department of Astronomy, The Ohio State University\\
    140 West 18th Ave, Columbus OH 43201, USA\\
    somers@astronomy.ohio-state.edu}

\def\teff   {{$T_{\rm eff}$}}
\def\teffs  {{$T_{\rm eff}s$}}
\def\msun   {{M$_{\odot}$}}
\def\ML     {{$\alpha_{\rm ML}$}}
\def\MLsol  {{$\alpha_{\odot}$}}
\def\delR   {{$\delta_{\rm R}$}}
\def\logg   {{$\log g$}}
\def\loggs  {{$\log g$s}}
\def\fspot  {{$f_{spot}$}}
\def\xspot  {{$x_{spot}$}}
\def\bf     {{$<$$B_* f_*$$>$}}

\begin{abstract}
We assess the impact of starspots on the evolution of late-type stars during the pre-main sequence (pre-MS) using a modified stellar evolution code. We find that heavily spotted models of mass 0.1-1.2\msun\ are inflated by up to $10$\% during the pre-MS, and up to 4\% and 9\% for fully- and partially-convective stars at the zero-age MS, consistent with measurements from active eclipsing binary systems. Spots similarly decrease stellar luminosity and \teff, causing isochrone-derived masses to be under-estimated by up to a factor of $2 \times$, and ages to be under-estimated by a factor of 2-10$\times$, at 3~Myr. Consequently, pre-MS clusters and their active stars are systematically older and more massive than often reported. Cluster ages derived with the lithium depletion boundary technique are erroneously young by $\sim 15$\% and $10$\% at $30$ and $100$ Myr respectively, if 50\% spotted stars are interpreted with un-spotted models. Finally, lithium depletion is suppressed in spotted stars with radiative cores, leading to a fixed-temperature lithium dispersion on the MS if a range of spot properties are present on the pre-MS. Such dispersions are large enough to explain Li abundance spreads seen in young open clusters, and imply a range of radii at fixed mass and age during the pre-MS Li burning epoch. By extension, this implies that mass, composition, and age do not uniquely specify the HR diagram location of pre-MS stars.
\end{abstract}

\section{Introduction}\label{sec1}

Precision measurements of fundamental stellar parameters active, late-type stars are systematically larger by $\sim$5-15\% than predictions from modern evolution codes, a phenomenon we refer to as radius inflation. This effect has been claimed in open clusters \citep[e.g.][]{jackson14a} and in field stars \citep[e.g.][]{popper97,torres02,ribas03,lopez-morales05,morales08,boyajian12}; on the main sequence (MS), and the pre-MS \citep[e.g.][]{stassun07}; and for stars above and below the fully-convective boundary \citep[e.g.][]{torres10,feiden12a}. The cause remains controversial, but a vital clue comes from the claimed correlation of radius inflation with magnetic activity \citep{torres06,lopez-morales07,morales08,stassun12}. Mechanisms proposed to induce this correlation include the magnetic inhibition of convection \citep[e.g.][]{chabrier07}, magnetic fields adding internal pressure support \citep[e.g.][]{mullan01}, and starspots \citep[e.g][]{torres06}, though each are likely at work to some degree in active stars. Although evidence is accruing that such radius anomalies are real, interpreters must be cautious as complications such as mis-estimated opacity \citep[e.g.][]{terrien12}, unaccounted systematics in interferometic measurements \citep{casagrande14}, and the impact of spots on radius measurements in eclipsing binaries \citep{morales10} may be mistaken for signatures of radius inflation.

In \citet[][SP14 hereafter]{somers14}, we found that radius inflation during the pre-MS has a strong impact lithium abundances of young stars. Radius inflation reduces the temperature at the base of the surface convection zone during the pre-MS, leading to a suppressed rate of Li destruction in standard stellar models. If stars of equal mass and age have different intrinsic sizes during the pre-MS, they will destroy different quantities of Li during the pre-MS, thus providing a mechanism for inducing the Li abundance dispersion observed in the K dwarfs of the Pleiades and other young clusters \citep[e.g.][]{soderblom93}. In \citet[][SP15 hereafter]{somers15}, we further showed that radius inflation can suppress Li destruction caused by rotational mixing on the pre-MS \citep[e.g.][]{pinsonneault90}. This implies that if rotation correlates with an increased radius on the pre-MS, as claimed in some works \citep[e.g.][]{littlefair11,cottaar14}, the Li-rotation correlation observed in the Pleiades could also be produced. This is important because rotational mixing is correlated with rotation rate, producing a trend of opposite sign relative to the data. Both of our previous papers were constructed as general investigations of the effect of radius inflation on Li abundances, conducted by varying the mixing length parameter \citep[e.g.][]{chabrier07}, and as such did not model a unique physical mechanism. In this paper, we expand upon this work by exploring a specific potential cause of radius inflation: starspots.

Spots are the visual manifestation of concentrated magnetic fields in the convective envelopes of late-type stars. These fields inhibit the flow of energy in specific regions near the surface, inducing a structural response \citep{gough66,spruit82a,spruit82b}. Spots have long been argued to affect the luminosities, temperatures and radii of stars \citep[e.g.][]{spruit82a,spruit86}, but in only a few cases have their effect on the pre-MS been examined \citep[e.g.][]{jackson14a}. \citet{spruit86} proposed a method for introducing the physical effects of spots into stellar evolution codes, with the goal of exploring the colors of stars with black spots at the zero-age MS (ZAMS). In this paper, we implement a similar treatment into our evolutionary code, and explore the impact of spottiness on pre-MS and ZAMS stellar parameters, the inference of masses and ages from the Hertzsprung-Russell (HR) diagram, the observed colors of stars with non-zero spot temperatures, and the destruction of lithium. By comparing our model predictions to data, a consistent picture emerges: pre-MS stars have a range of radii at fixed mass and age, correlated with rotation, which persists at the ZAMS.

The text is organized as follows. We discuss our evolutionary models in $\S$\ref{sec2}, with a particular emphasis on our treatment of spots ($\S$\ref{sec2.2}). With our models calculated, we first consider the impact of spots during the pre-MS and at the ZAMS on stellar radii, luminosities, and \teffs\ in the mass range 0.1-1.2\msun\ ($\S$\ref{sec3}). Next, we discuss how spots might introduce errors into the quantification of masses and ages using isochrones ($\S$\ref{sec4.1}), and how spots affect observed colors ($\S$\ref{sec4.2}). Then, we examine how spots impact the destruction of lithium, both in the context of age determinations with the lithium depletion boundary technique, and in the context of Li abundance spreads observed in young MS clusters ($\S$\ref{sec5}). In $\S$\ref{sec6}, we synthesize the results of $\S$\ref{sec3}$-$\ref{sec5}, compare our quantitative findings to other work in the literature, and discuss some outstanding questions for starspot modeling. Finally, we summarize our conclusions in $\S$\ref{sec7}.

\section{Methods}\label{sec2}

\subsection{Standard stellar models}\label{sec2.1}

We calculate non-rotating, solar-metallicity stellar models with the Yale Rotating Evolution Code (YREC; e.g. \citealt{vansaders13}). For the purposes of calibration, we assume the Sun is unaffected by the presence of spots; this is reasonable, given the Sun's minimal spot coverage. We adopt a present day solar heavy element abundance of $Z/X = 0.02292$ from \citet{gs98}, and choose the hydrogen mass fraction ($X$) and the mixing length coefficient (\ML) that reproduce the solar luminosity and radius for a 1\msun\ model at $4.568$~Gyr. The final solar calibrated values are $X = 0.71172$, $Z = 0.018923$, and \MLsol\ $= 1.92531$. Our models use the 2006 OPAL equation of state \citep{rogers96,rogers02}, model atmospheres from \citet{allard97}, high temperature opacities from the opacity project \citep{mendoza07}, low temperature opacities from \citet{ferguson05}, and nuclear cross-sections from \citet{adelberger11}. We treat electron screening using the method of \citet{salpeter54}. 

Our models are initialized above the Hayashi track, and evolved to the age at which 99\% of the initial deuterium abundance has been destroyed \citep[e.g.][]{stahler88}. We define this time as $t = 0$ for each individual model. This approach represents a physical scenario: as proto-stars contract and build up mass through accretion, their interiors increase in temperature. Eventually, the core becomes hot enough to fuse deuterium, a process which partially supports the star against gravitational contraction; this is the so-called ``deuterium birth line''. Proto-stars travel along this sequence towards higher mass as accretion continues to supply fresh deuterium fuel to the central engine. Once the accretion rate subsides, the star will begin to contract towards the hydrogen burning main sequence. Our zero point provides a consistent criteria for the starting age of all models, and ensures that models of equal mass but differing spot properties are in equivalent evolutionary states when compared at early ages.

\subsection{Spots}\label{sec2.2}

\subsubsection{Implementation}

Our spot methodology, based on the treatment by \citet{spruit86}, incorporates two effects: the inhomogeneous photosphere in the presence of spots, and the inhibition of convection in magnetic regions. To simulate this behavior, we divide our stellar models into two zones: un-spotted and spotted. We account for the flux redistribution by explicitly altering the radiative transport gradient in the surface convection zones of our models, and the inhomogeneous surface by properly treating the outer boundary condition (see below).

The spots in our models are controlled by two parameters: the filling factor \fspot, defined as the ratio of the spotted surface to the total surface area, and the spot temperature contrast ratio \xspot, defined as the temperature of the spotted surface divided by the temperature of the un-spotted photosphere. The surface-weighted average of the two zone temperatures defines the temperature of the model at a given radius.

\begin{equation} \label{eqn1}
T^4 = (1-f_{spot}) T_{amb}^4 + f_{spot} T_{spot}^4,
\end{equation}

$T_{spot}$ is the area-averaged temperature of the prenumbral and umbral spot regions, and $T_{amb}$ is the temperature of the ambient, un-spotted region. By substituting $T_{spot} = x_{spot} T_{amb}$, we find an expression for the total luminosity,

\begin{equation}\label{eqn2}
L = 4 \pi R^2 \sigma_{SB} T_{amb}^4 (1-f_{spot} + f_{spot} x_{spot}^4),
\end{equation}

and the flux in the ambient regions,

\begin{equation}\label{eqn3}
F_{amb} = F_{tot}/(1-f_{spot} + f_{spot} x_{spot}^4).
\end{equation}

This formulation implies a modeling degeneracy between \fspot\ and \xspot, since the flux redistribution is actually controlled by the redistribution parameter,

\begin{equation}\label{eqn4}
\alpha_{spot} = (1-f_{spot} + f_{spot} x_{spot}^4).
\end{equation}

$\alpha_{spot}$ can be thought of as one minus the effective blocking area of totally black (\xspot\ = 0) spots, a parameter typically called $\beta$ ($= 1 - \alpha_{spot}$). Eq. \ref{eqn3} governs the flux enhancement in the non-magnetic regions of the model. We implement this enhancement into our evolution code by dividing the radiative gradient in convective regions by $\alpha_{spot}$,

\begin{equation}\label{eqn5}
\nabla_{rad,amb} = \nabla_{rad} / \alpha_{spot}.
\end{equation}

Since \fspot\ and \xspot\ fall in the range 0--1 by definition, $\alpha_{spot}$ is always less than or equal to one, meaning that spots can only increase the radiative flux in the un-spotted regions. The luminosity and effective temperature of the star can be found by solving Eqs. \ref{eqn1} and \ref{eqn2} using the values of $T_{amb}$, $T_{spot}$, and \xspot\ at the surface where the optical depth $\tau = 2/3$. To model the changing temperature of spotted regions with depth, we hold the difference in temperature between the two zones fixed ($T_{amb} - T_{spot} = {\rm const.}$). The result is that spots are only significant near the surface, where $T$ is low. We test the importance of spot depth on our models in $\S$\ref{sec6.3.1}, and find that assumptions about the subsurface morphology impact radius, luminosity, and \teff\ predictions by $\sim 1$~\% at most.

Our spot treatment involves a crucial implicit assumption: the spotted and un-spotted atmospheric regions are in pressure equilibrium. This means that if the spotted regions contain a magnetic field which produces a pressure $P_{spot,mag}$, the gas pressure in the two regions will be such that $P_{amb,gas} = P_{spot,gas} + P_{spot,mag}$\footnote{A further assumption of magnetic and gas equiparition therefore implies $P_{amb,gas} = 2 \times P_{spot,gas}$, a fact we make use of in $\S$\ref{sec6.2}}. This requirement permits us to obtain the surface pressure in the un-spotted region, and apply it equally in both zones. $P_{surf}$ is obtained in our evolution code by using a model atmosphere, which returns a pressure $P_{surf}$ for a given \teff\ and $\log g$. For consistency, we call our atmosphere routine using the temperature of the un-spotted, ambient surface:

\begin{equation}\label{eqn6}
P_{surf} = P_{surf}(T_{amb,surf},\log g).
\end{equation}

Our approach differs from previous studies which have considered the effects of magnetic activity on stellar properties. \citet{spruit86} used a similar spot formulation to investigate the impact on stellar parameters, but focused primarily on the ZAMS, and did not consider the effect of non-zero spot temperatures on the location of stars in the color-magnitude diagram. In a recent series of papers, \citet{feiden12b} examined the consequences of magnetic fields in stellar interiors by directly incorporating a 1-dimensional approximation of the induction equation into their model physics \citep{feiden13,feiden14}. This approach differs from ours by considering a one-zone, homogeneous surface boundary condition, compared to the two-zone, inhomogeneous approach we have adopted. However, since magnetic fields are the underlying cause of surface spots, our work compliments theirs by approaching the same mechanism from a different perspective. In addition, \citet{jackson14a,jackson14b} have recently examined the impact of surface spots on the pre-MS radii of late-type stars, and the lithium destruction rates of stars with M $<$ 0.5\msun. Their procedure relied on calibrating polytropic models on published evolutionary tracks, and then considering the impact of flux-blocking spots at the surface. By contrast, we employ a self consistent evolutionary code to produce our calculations, and thus anticipate somewhat different results ($\S$\ref{sec6.2}). Each of these studies differed in their treatment from our present study, and we consider our results complimentary to theirs.

\subsubsection{Adopted spot properties}

In this section, we select a range of spot properties to investigate in our models. We are interested in the range of stellar properties that could be produced by spots, as well as their dependence on evolutionary state and mass. Most measured filling factors to date have been for dwarfs, sub-giants, and giants (see tab. 9 in \citealt{berdyugina05}). While very few pre-MS stars have similar measurements, perhaps the most direct comparison are the spotted, tidally-synchronized sub-giants known as RS CVn stars. RS CVns are structurally similar to pre-MS stars in the outer regions, as they have deep convective envelopes and low gravities relative to the MS, and thus host similar environments for spot emergence. Claimed filling factor among this class reach $40$ to $50$\%, such as in the cases of HU Vir, UX Ari, and HR1099 \citep{o'neal98,o'neal01}, demonstrating that deep envelopes paired with moderate ($5-10$~d) rotation periods can produce a high level of spot coverage. The T Tauri star V410 Tau has a measured surface spot coverage of 32-41\% \citep{petrov94}, supporting the notion that pre-MS stars can be similarly spotted. Furthermore, Zeeman-Doppler imaging has confirmed average surface fields of up to a few kG on pre-MS stars \citep[e.g.][]{reiners12}, comparable to the surface fields of highly active, heavily spotted MS stars (see below). Collectively, these arguments support the hypothesis that pre-MS stars can host spots which approach 50\% coverage.

Similar filling factors have been claimed for young MS stars. For example, the rapidly rotating ($P_{\rm rot} \sim 1$ to $4$~d) dwarfs LQ Hya, V833 Tau, and EQ Ver host filling factors of $40$ to $50$~\% \citep{o'neal01,o'neal04}. It may appear surprising at first that the slower rotating RS CVn stars (e.g. $P_{\rm rot} \sim 10.4$~d for HU Vir; \citealt{o'neal98}) can be as spotted as the faster rotating ZAMS stars. However, the convective overturn timescale of sub-giants is much longer than MS dwarfs, implying a smaller Rossby number (ratio of rotation rate to overturn timescale) at fixed rotation period. Magnetic properties appear to scale more closely with Rossby number than with absolute rotation \citep[e.g.][]{noyes84,pizzolato03}, so it is reasonable to expect equivalent spot properties at slower rotation rates for sub-giants. This argument applies equally well to pre-MS stars, whose average rotation rates are slower at 3 Myr than at the ZAMS \citep[e.g.][]{littlefair10}, but whose convection zones are deeper.

We conclude that \fspot\ = 0.5 is a reasonable assumption for the upper limit of pre-MS and ZAMS filling factors, and we adopt this. We further adopt $x_{spot} = 0.8$, characteristic of the range of observed spot temperature ratios (0.65 to 0.85; e.g. \citealt{berdyugina05}). A fuller treatment would include a mapping of spot temperature and filling factor as a function of age, metallicity, mass, and rotation rate; the current exercise is aimed at a different purpose, namely constraining the potential impact of star spots on observables. We do note that the Pleiades lithium dispersion (see $\S$\ref{sec5}) does provide evidence for a range of spot properties in pre-MS stars, as we argued in \citetalias{somers15}.

\section{Effect on stellar radii}\label{sec3}

\subsection{The pre-main sequence}\label{sec3.1}

\begin{figure}
\includegraphics[width=3.3in]{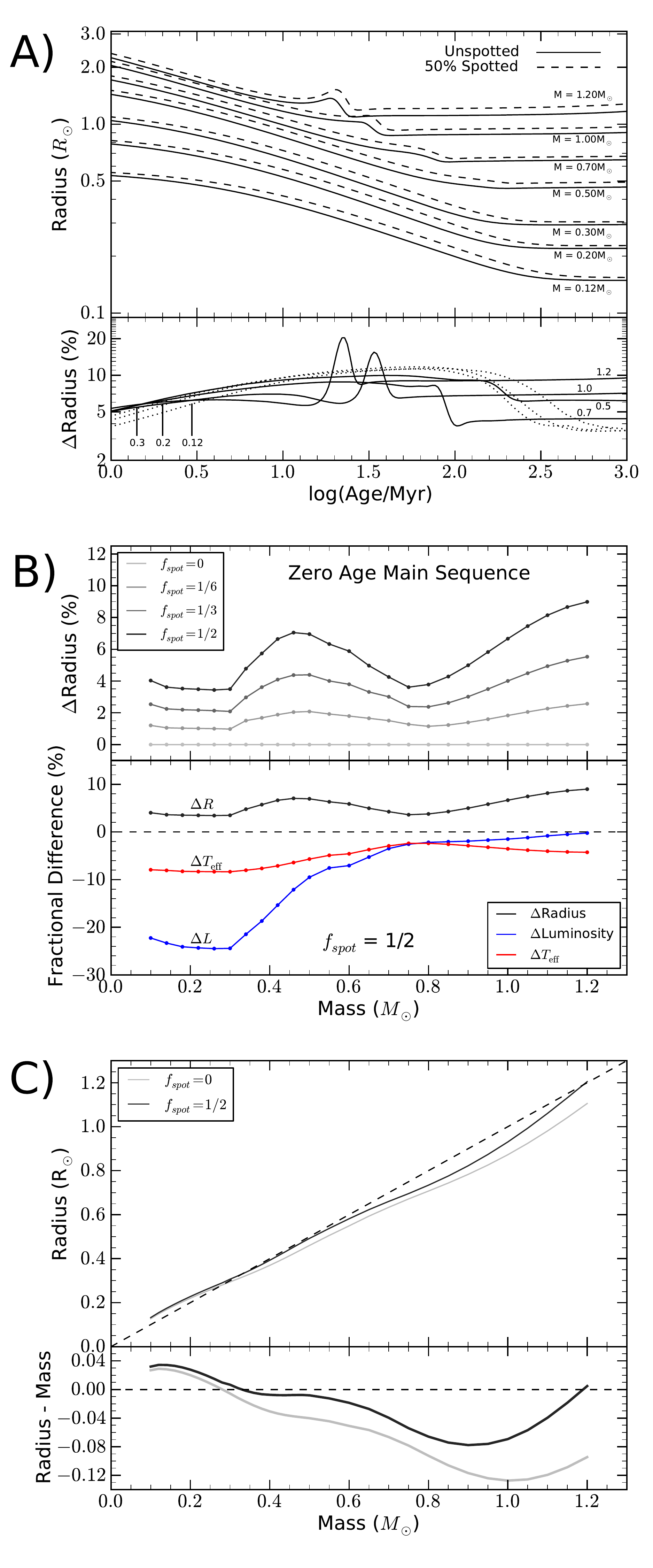}
\caption{The impact of spots on radii, luminosities, and \teffs\ of stars in the mass range 0.1-1.2\msun. \textit{A)} The time evolution of the radius of un-spotted (solid) and spotted (dashed) models of various masses are shown in the top. The bottom shows evolution of the difference in radius between the spotted and un-spotted models for fully convective (dotted) and partially convective (solid) models. $\Delta R$ steadily grows during the early pre-MS, before stabilizing at the entrance to the MS. \textit{B)} The top shows zero-age MS radius anomalies across our mass range, for three different values of \fspot. The bottom plots the luminosity and \teff\ anomalies resulting from the case of \fspot\ = 0.5, respectively shown in blue and red. \textit{C)} The mass-radius relationship for un-spotted (light) and spotted (dark) models. The bottom panel shows that the mass-dependence of the structure of the relationship is affected by spots; this leads to the strong mass dependence of $\Delta R$ shown in panel B.}
\label{fig1}
\end{figure}

We first explore the impact of spots on stellar model radii during the pre-MS. Evolutionary tracks representing both fully convective stars ($M \lesssim 0.35$\msun) and stars which develop radiative cores ($0.35$\msun $\lesssim M \lesssim 1.2$\msun) are shown in the upper panel of frame A in Fig. \ref{fig1}. Solid black lines represent the radius as a function of time for un-spotted models, and the dashed black lines represent the same for models half covered in 80\% temperature contrast spots (\fspot\ = 0.5, \xspot\ = 0.8, $\alpha_{spot} \sim 0.7$), referred to hereafter as our base case. The difference in radius between equal mass, un-spotted and spotted models as a function of age is reported in the lower panel.

Spots lead to an inflated radius at all ages, but the magnitude evolves during the pre-MS. For each mass, the radius enhancement gradually increases with age, initially peaking around 8-12\%, and finally stabilizing on the MS. The age dependence of pre-MS radius inflation results primarily from a timing argument: the rate of pre-MS contraction is proportional to stellar luminosity, and because un-spotted stars are more luminous than their spotted counterparts on the pre-MS (see below), they will have contracted further at a given age. This explains the gradual increase in $\Delta R$ with time, and leads to a conspicuous transient spike in $\Delta R$ at the onset of MS burning, as the un-spotted star has reached the ZAMS while the spotted star remains on the late pre-MS. However the internal structure is also altered in a mass dependent fashion: changes to the efficiency of convection impact higher mass models \citep[e.g.][]{andronov04}, while changes to the surface boundary condition are important at lower mass \citep[e.g.][]{spruit86}.

Spotted models show corresponding anomalies in both \teff\ and luminosity. The average \teff\ for our base case is lower by up to 7\% for all masses, and corresponds to a zero point shift in the Hayashi track. Pre-MS luminosity anomalies therefore evolve according to $L/R^2 \sim \rm{const}$. Luminosities of the spotted models are lower by up to 18\% and 10\% during the early and late pre-MS for high mass stars, and lower by up to 25\% and 10\% during the early and late pre-MS for low mass stars. We explore consequences of these \teff\ and $L$ anomalies in $\S$\ref{sec4} and $\S$\ref{sec5}.

\subsection{The zero-age main sequence}\label{sec3.2}

Once at the ZAMS, radius anomalies converge to a stable, mass-dependent pattern. The black line in the upper panel of frame B in Fig. \ref{fig1} shows our base case radius anomalies at the age when 0.1\% of the core hydrogen has been burned. As can be seen, fully convective stars have a weakly mass-dependent anomaly of $\sim$4\%, and higher mass stars generally have larger inflation factors with local maxima at 0.5\msun\ and 1.2\msun, and a local minimum at 0.7\msun. Anomalies resulting from different values of \fspot\ are also shown. Unsurprisingly, the radius enhancement increases self-similarly with greater spot coverage, and the change in $\Delta R$ with \fspot\ is approximately linear. These results are generally consistent with \citet{spruit86}, though the second minimum around 0.8\msun\ is not present in their ZAMS models. This may be due to differences in the adopted standard model physics.

\begin{figure*}
\includegraphics[width=7.0in]{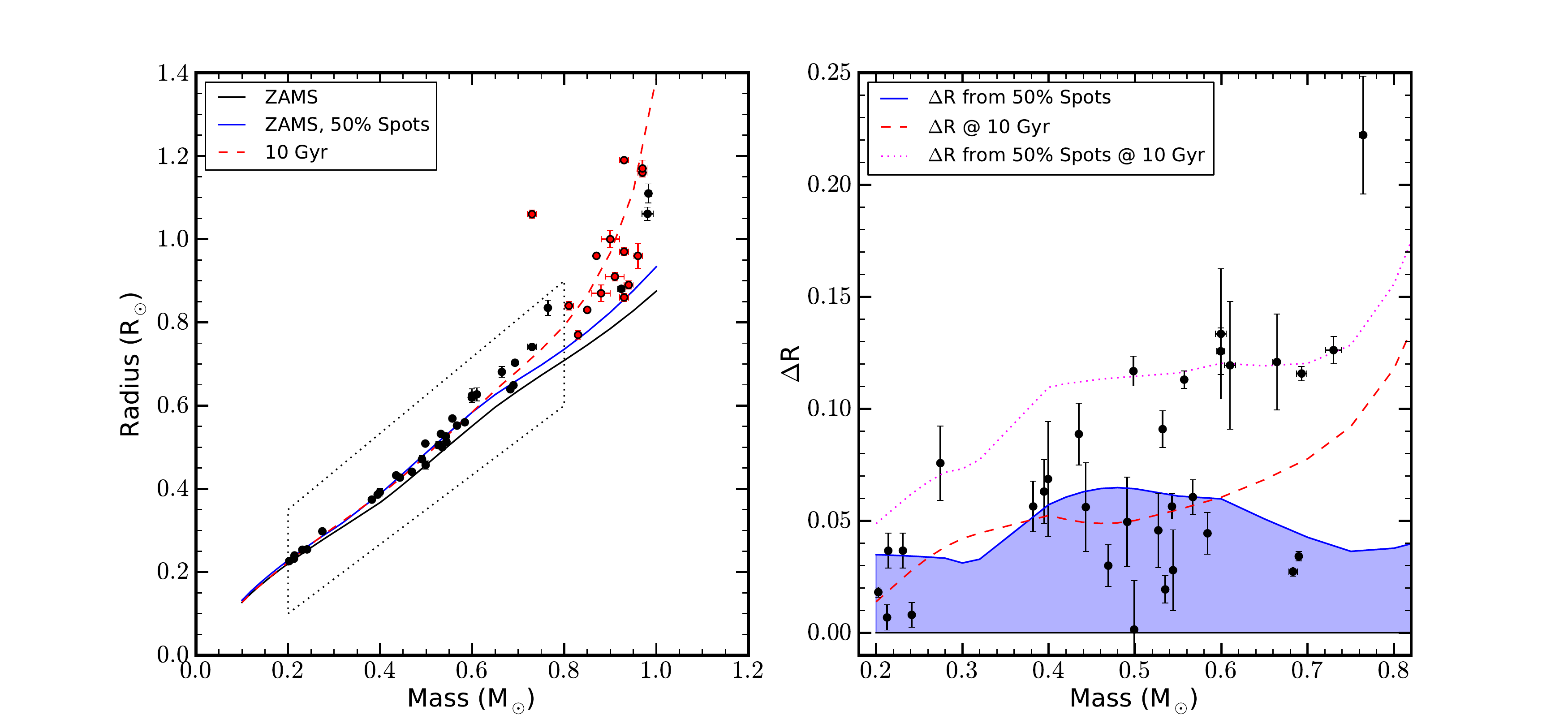}
\caption{\textit{Left:} The mass-radius relationship for un-spotted (black) and spotted (blue) models at the ZAMS, and the 10 Gyr relation for un-spotted models (red dashed). These are compared to data from eclipsing binaries (see the text for references). \textit{Right:} The same as the left, but detrended against the ZAMS, un-spotted isochrone. Dotted purple represents the sum of age and spot effects. Two-thirds of the data lie within the spotted models, and all but one lie within the combined effects of activity and age.}
\label{fig2}
\end{figure*}

This radius enhancement in spotted models is predominately a result of the altered boundary conditions at the surface. The photospheric pressure is higher for the hotter ambient photosphere than it would be for a model with $T_{surf} =$~\teff\ at fixed radius. The model expands as a result in order to reach equilibrium ($P \propto M^2 / R^4$ in homology), which lies at higher radius and lower surface pressure. This reduced $P$ leads to a corresponding decrease in $P$ throughout the surface convective zone, due to its adiabatic stratification. For fully convective stars, the pressure reduction reaches the nuclear burning core where it directly suppresses the rate of reactions, and thus lowers the luminosity. The blue line in the lower panel of frame B shows $\Delta L$ as a function of mass, revealing a significant decrease in the luminosity of fully convective stars due to our base case spots. This effect is less pronounced in higher mass stars, whose convection zones have receded from the core and therefore exert less influence on the central engine \citep{kippenhahn90}. The decreasing luminosity anomaly from 0.3--0.6\msun\ leads to a spike in the radius and temperature (red line in frame B) anomalies, which peak around 0.5\msun\ and 0.7\msun\ respectively. Note that above $\sim$0.5\msun, $\Delta R \approx -2 \Delta T$, in good agreement with observational studies \citep[e.g.][]{morales08}.

The peculiar structure of the dependence of $\Delta R$ on mass can be understood by considering the spotted and un-spotted mass-radius relationships plotted in the upper panel of frame C in Fig. \ref{fig1}. As can be seen, between 0.1 and 1.0\msun\ the slope of the mass-radius relation at first decreases, then increases, decreases, and increases again. These ``wiggles" result from mass-dependent changes in stellar structure and energy generation, such as the onset of a radiative core, the decreasing mass of the convective envelope, and the burgeoning dominance of CNO hydrogen processing. The lower panel shows both mass-radius relations detrended with the line $M = R$ (in solar units). The inclusion of spots alters the masses at which these transitions occur, shifting the onset of slope changes to lower mass, and thus leading to significant structure in the mass dependence of radius inflation. 

\subsection{Comparison with data}\label{sec3.3}

In this section, we compare the mass-radius relationship predicted by our models to literature data. The left panel of Fig. \ref{fig2} displays our predicted relationships for un-spotted (black line) and spotted (blue line) models at the ZAMS, and for un-spotted models at 10 Gyr (dashed red). These are compared to $M \leq 1.0$\msun\ eclipsing binary data from \citet{torres10} and \citet{feiden12a} (red and black, respectively). The binaries generally form a tight sequence bracketing the spotted isochrone below 0.8\msun, and fan out into a larger range above 0.8\msun, attesting the importance of age effects at a solar mass and above. In the right panel, the data falling within the dotted parallelogram are detrended with the standard ZAMS isochrone, and compared to the spotted and 10 Gyr isochrones. The data occupy an envelope $\sim4-7$\% at the low mass end, and increase to $\sim$12\% at 0.7\msun, demonstrating the inflated radius problem discussed in $\S$\ref{sec1}. Two-thirds of the detrended stars fall within the shaded blue region, and all but one fall within the combined effects of age and spots\footnote{Old stars are normally not expected to be heavily spotted, but short period eclipsing binaries are often tidally locked, permitting rapid rotation at late ages.} (dotted purple line). This shows that the radius anomalies resulting from realistic spot parameters in our models are consistent with empirical constraints on radius inflation. 

\begin{figure*}
\includegraphics[width=7.0in]{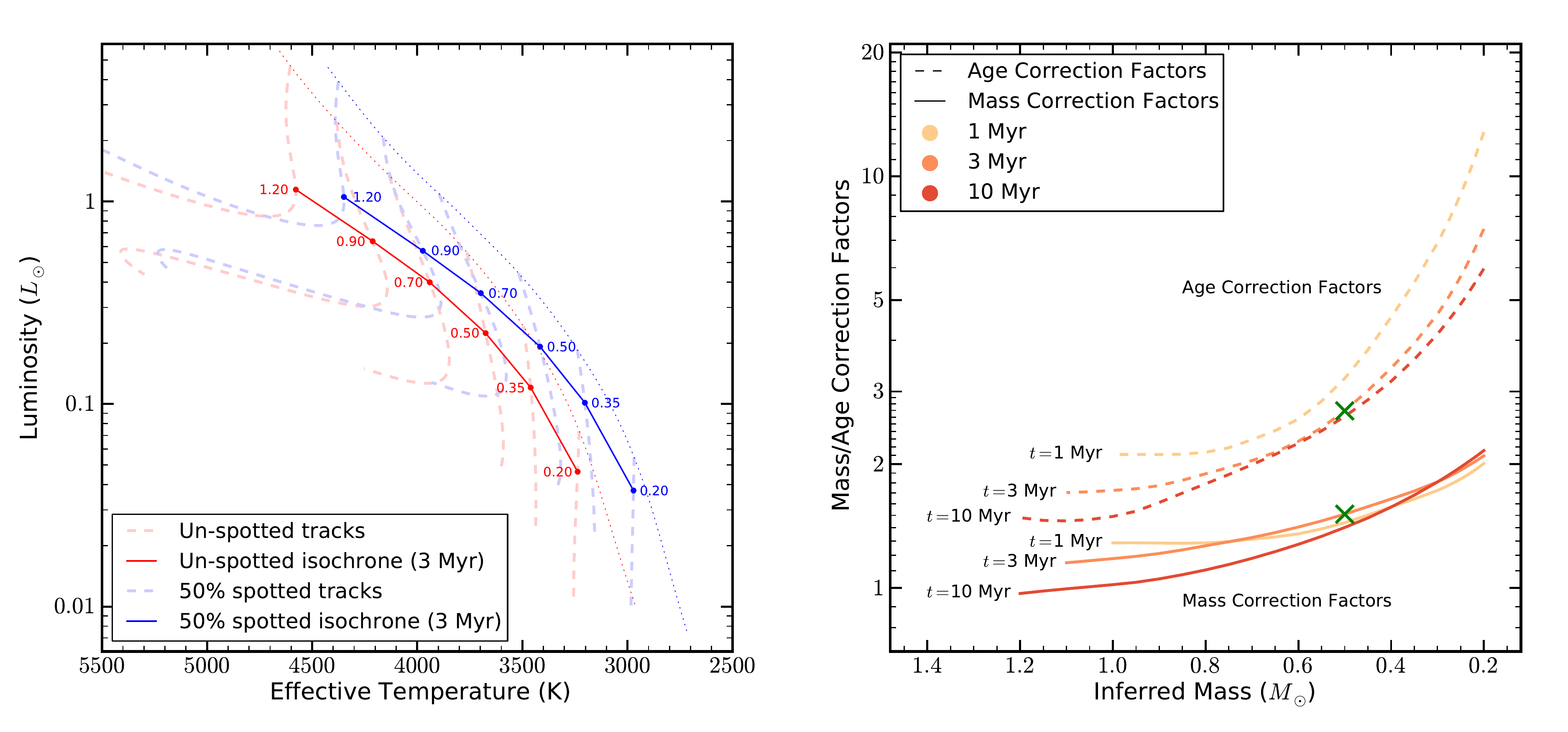}
\caption{The impact of spots on the HR diagram. \textit{Left:} Un-spotted (red) and spotted (blue) evolutionary tracks and 3 Myr isochrones for a range of masses. Spots displace stars to the lower right (cooler and less luminous) at fixed age. As a result, a dispersion in luminosity at fixed \teff\ develops across the mass function, and spotted stars appear less massive and younger when interpreted with un-spotted isochrones. \textit{Right:} Correction factors which describe the errors incurred when measuring the mass and age of spotted stars with un-spotted isochrones. At a given mass (x-axis) and age (red/orange/yellow) inferred from un-spotted isochrones, the y-axis reports the necessary corrections. For example, a heavily spotted star which appears to have a mass of 0.5\msun\ and an age of 3 Myr will have respective corrections (green crosses) of 1.5 (implying a true mass of $1.5 \times 0.5$~\msun\ $\sim 0.75$~\msun), and 2.7 (implying a true age of $2.7 \times 3$~Myr $\sim 8$~Myr).}
\label{fig3}
\end{figure*}

We find no clear evidence for higher average anomalies around 0.5\msun\ compared to 0.75\msun\ (see $\S$\ref{sec3.2}), but a host of systematic issues forbid any present conclusions. For example, stars of equal mass but unequal metal abundance will have differing radii, but we find that this dispersion is only $\sim 2$\% for [Fe/H]~$\in$~$[-0.3,+0.3]$. Additionally, surface spots can increase the inferred transit time for eclipsing binaries, thus leading to systematic over-estimates of up to $\sim$3\% in the radius \citep[e.g.][]{morales10}. Finally, age effects will contribute to inferred anomalies, as suggested by Fig. \ref{fig2}. \citet{feiden12a} have recently argued that accounting for these effects reduces most radius anomalies to around 5\%, instead of the often quoted range of 5-15\%, in good agreement with our expectations. In any case, it appears that spot-induced radius effects are of a similar magnitude as the well-known radius anomalies in eclipsing binaries, and should not be neglected when interpreting these observations.

\section{Effect on inferred masses, ages, and colors}\label{sec4}

Measuring the masses and ages of pre-MS stars is a challenging endeavor, given the rapid evolutionary timescales and a host of observational issues \citep[e.g.][]{hartmann01}, but is of great importance for studies of the initial mass function, circum-stellar discs, the early angular momentum evolution of stars, and the formation of planets. The ideal approach would be to compare calibrated pre-MS isochrones with the HR diagram of a cluster, and to read off the masses and ages of its members. In reality, this procedure is complicated by a variety of observational and theoretical issues, which broadly fall into three categories.

\textit{Observational errors.} In young clusters, extinction is highly variable from source to source, binarity status is often unknown, and stellar magnitudes may vary on short and long timescales due to rotation, disc obscuration, and accretion. These effects form a base level of uncertainty.

\textit{Theoretical issues.} Pre-MS evolutionary tracks lack a good calibration data set, and different suites make varying predictions for the birth line location, the bolometric corrections, and the rate of Hayashi contraction. This uncertainty severely limits one's ability to derive masses and ages in young clusters.

\textit{Non-uniform parameters.} Even if these issues were resolved, the possibility remains that stellar parameters such as luminosity and radius vary between cluster members at fixed mass and age, perhaps due to physics associated with rotation, magnetic activity, or accretion. 

The picture is further complicated by intra-cluster age spreads, which mimic the appearance of non-uniform stellar parameters by producing luminosity dispersions at fixed \teff. In order to separate these effects, it is important to asses the impact that non-uniform parameters would have on coeval stars.

In $\S$\ref{sec3}, we showed that spots induce shifts to the radii, luminosities and \teffs\ of stars during the pre-MS, and can therefore generate non-uniform parameters among young cluster members. In this section, we explore the impact of these shifts on the Hertzprung-Russell and color-magnitude diagrams of young clusters, and discuss their implications for understanding pre-MS stars.

\subsection{Spot effects in the Hertzsprung-Russell diagram}\label{sec4.1}

We first constrain ourselves to the theoretical HR diagram, and consider the impact of spots on the temperature and luminosity of our stellar models. This for now ignores the impediments to obtaining these parameters brought on by spots (see $\S$\ref{sec4.2}), and assumes the true surface \teff\ and $L$ are known.

The red dashed lines in the left panel of Fig. \ref{fig3} show un-spotted tracks for a range of masses, run from the deuterium birth line (dotted red line) to 50 Myr. The blue dashed lines represent the same for our base case spotted models. Isochrones of 3 Myr are over-plotted in solid red and blue, with masses periodically enumerated. The results of $\S$\ref{sec3} are reflected in this figure: spots reduce the intrinsic luminosity and \teff\ of stars in a mass-dependent fashion. The fractional \teff\ displacement ranges from $\sim$6--8\% from the highest to the lowest mass, which corresponds to a nearly mass-independent decrement of 270~K. The fractional luminosity reduction ranges from $\sim$10--22\% from high to low mass, corresponding to shifts of 0.1 -- 0.01 $L_{\odot}$. A range of spot filling factors will therefore induce a significant dispersion in $L$ at fixed \teff. However, even in our extreme base case the dispersion does not reach the order-of-magnitude spread observed in some young clusters \citep[e.g.][]{hillenbrand97}. Thus spots may be an important contributor to young cluster dispersions, but not their sole cause. We emphasize that these shifts are not an observational bias, but \textit{a change in the true temperature and luminosity of pre-MS stars.}

{\scriptsize
\setlength{\tabcolsep}{1.4pt}
\renewcommand{\arraystretch}{0.55}

\begin{table}[!*hb]
\caption{Mass and Age Correction Factors}
\label{tab:factors}
\begin{tabular}{ccccccccc}

\hline \hline
Inferred Mass&&\multicolumn{3}{c}{Mass Correction}&&\multicolumn{3}{c}{Age Correction} \\
(\msun)&&\multicolumn{3}{c}{Factors}&&\multicolumn{3}{c}{Factors} \\
\hline \\

\underline{\textbf{\fspot\ = 1/6}}&&{\textbf{1 Myr}}&{\textbf{3 Myr}}&{\textbf{10 Myr}}&&{\textbf{1 Myr}}&{\textbf{3 Myr}}&{\textbf{10 Myr}} \\ 
\cline{3-5} \cline{7-9} \\
0.20&&1.263&1.275&1.296&&3.589&2.311&1.881 \\
0.22&&1.249&1.263&1.289&&3.280&2.191&1.835\\
0.24&&1.234&1.251&1.281&&3.010&2.084&1.790\\
0.26&&1.220&1.239&1.270&&2.777&1.986&1.740\\
0.28&&1.205&1.226&1.263&&2.561&1.893&1.704\\
0.30&&1.194&1.214&1.255&&2.397&1.818&1.668\\
0.35&&1.175&1.194&1.236&&2.120&1.686&1.597\\
0.40&&1.163&1.181&1.217&&1.939&1.600&1.535\\
0.45&&1.151&1.172&1.198&&1.782&1.532&1.495\\
0.50&&1.137&1.160&1.178&&1.648&1.464&1.449\\
0.55&&1.123&1.148&1.155&&1.535&1.407&1.406\\
0.60&&1.106&1.135&1.133&&1.434&1.357&1.368\\
0.65&&1.095&1.119&1.111&&1.383&1.312&1.332\\
0.70&&1.093&1.109&1.090&&1.360&1.286&1.295\\
0.80&&1.093&1.097&1.059&&1.335&1.264&1.254\\
0.90&&1.092&1.085&1.036&&1.313&1.242&1.196\\
1.00&&1.089&1.074&1.013&&1.294&1.226&1.163\\
1.10&&1.088&1.063&1.002&&1.286&1.208&1.135\\
1.20&&...&1.052&0.994&&...&1.193&1.127\\ \\

\underline{\textbf{\fspot\ = 1/3}}&&{\textbf{1 Myr}}&{\textbf{3 Myr}}&{\textbf{10 Myr}}&&{\textbf{1 Myr}}&{\textbf{3 Myr}}&{\textbf{10 Myr}} \\ 
\cline{3-5} \cline{7-9} \\
0.20&&1.593&1.633&1.710&&7.349&4.321&3.419\\
0.22&&1.548&1.595&1.681&&6.412&3.937&3.244\\
0.24&&1.510&1.560&1.650&&5.686&3.622&3.083\\
0.26&&1.477&1.529&1.620&&5.104&3.358&2.938\\
0.28&&1.450&1.501&1.592&&4.641&3.139&2.813\\
0.30&&1.427&1.477&1.564&&4.268&2.958&2.702\\
0.35&&1.388&1.435&1.499&&3.606&2.635&2.487\\
0.40&&1.355&1.401&1.434&&3.114&2.399&2.291\\
0.45&&1.321&1.370&1.375&&2.714&2.204&2.132\\
0.50&&1.284&1.338&1.321&&2.383&2.034&1.999\\
0.55&&1.248&1.303&1.271&&2.125&1.887&1.883\\
0.60&&1.220&1.271&1.224&&1.949&1.772&1.781\\
0.65&&1.206&1.243&1.184&&1.856&1.689&1.694\\
0.70&&1.202&1.223&1.148&&1.801&1.639&1.627\\
0.80&&1.195&1.191&1.092&&1.723&1.571&1.524\\
0.90&&1.183&1.161&1.046&&1.660&1.513&1.403\\
1.00&&1.182&1.132&1.017&&1.644&1.460&1.327\\
1.10&&1.182&1.110&1.000&&1.640&1.428&1.280\\
1.20&&...&...&0.984&&...&...&1.279\\ \\

\underline{\textbf{\fspot\ = 1/2}}&&{\textbf{1 Myr}}&{\textbf{3 Myr}}&{\textbf{10 Myr}}&&{\textbf{1 Myr}}&{\textbf{3 Myr}}&{\textbf{10 Myr}} \\ 
\cline{3-5} \cline{7-9} \\
0.20&&2.005&2.093&2.156&&12.819&7.453&5.968\\
0.22&&1.931&2.023&2.074&&11.039&6.649&5.490\\
0.24&&1.866&1.961&2.000&&9.639&5.993&5.076\\
0.26&&1.812&1.904&1.931&&8.526&5.453&4.721\\
0.28&&1.765&1.854&1.867&&7.630&5.003&4.413\\
0.30&&1.725&1.809&1.808&&6.888&4.627&4.139\\
0.35&&1.643&1.719&1.680&&5.523&3.928&3.591\\
0.40&&1.573&1.644&1.571&&4.545&3.419&3.177\\
0.45&&1.502&1.575&1.478&&3.789&3.017&2.864\\
0.50&&1.439&1.510&1.401&&3.229&2.692&2.609\\
0.55&&1.390&1.454&1.333&&2.841&2.450&2.407\\
0.60&&1.352&1.405&1.275&&2.571&2.270&2.245\\
0.65&&1.329&1.362&1.224&&2.409&2.134&2.110\\
0.70&&1.313&1.325&1.179&&2.291&2.038&1.992\\
0.80&&1.286&1.265&1.105&&2.134&1.893&1.790\\
0.90&&1.285&1.211&1.051&&2.108&1.772&1.612\\
1.00&&1.286&1.177&1.018&&2.108&1.725&1.491\\
1.10&&...&1.150&0.994&&...&1.703&1.454\\
1.20&&...&...&0.968&&...&...&1.480\\ \\

\end{tabular}

\tablecomments{Mass and age correction factors for three spot filling factors at various inferred masses and ages. Correct useage is detailed in $\S$\ref{sec4.1}. Ellipses indicate inferred mass and age points outside of the interpolation grid.}

\end{table}
}

Pre-MS Hayashi evolution proceeds at approximately fixed \teff, with $L$ declining with age. Our models demonstrate this behavior, but also reveal an important result: spots alter the \teff\ at which this evolution occurs. Spots displace stars downward and to the right in the plane of the HR diagram, and consequently spotted stars appear \textit{younger} and \textit{less massive} when interpreted with standard isochrones. To quantify these errors, we use the un-spotted evolutionary tracks of Fig. \ref{fig3} to infer the masses and ages of a range base case spotted models, and derive a set of correction factors to the resulting parameters. The results appear in the right panel of Fig. \ref{fig3}. These factors operate in the following sense: if the mass and age of a 50\% spotted star are derived from its HR diagram location using un-spotted tracks, the true mass and age will be given by the derived values multiplied by their respective correction factors (see the caption for a specific example). Factors for a variety of masses, ages, and filling factors can be found in Table \ref{tab:factors}. 

\begin{figure*}
\begin{centering}
\includegraphics[width=5.5in]{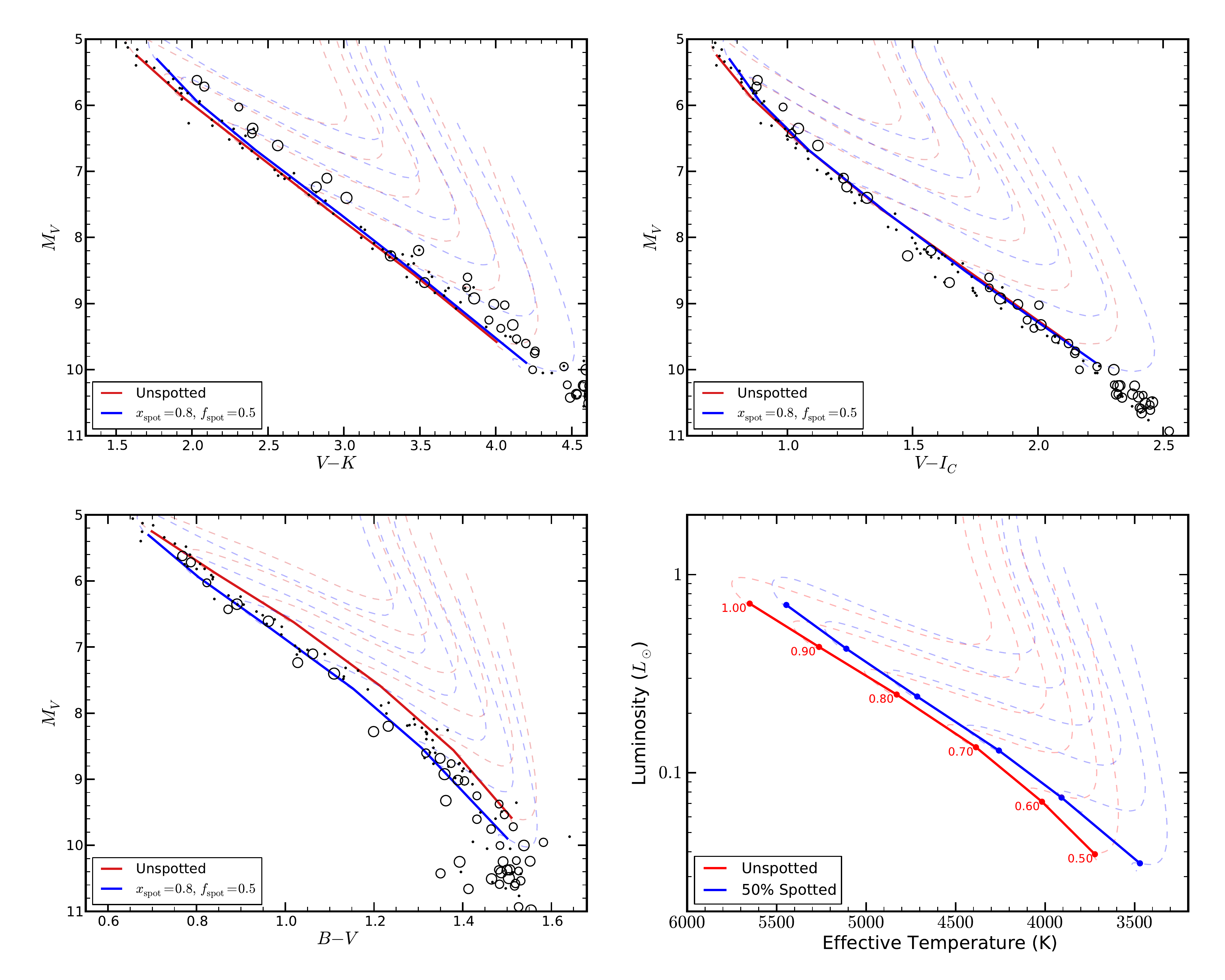}
\caption{The impact of spots on the color-magnitude diagram. Using empirical bolometric corrections, our un-spotted (red) models are directly transformed into three different color coordinates using a two-zone spot model. In color coordinates, the spotted models do not appear as cool due to the contribution to short wavelength emission from the un-spotted photosphere. The spot models appear bluer when shorter wave-lengths are used, due to the progressively stronger sampling of the Wein tail of the blackbody function. The bottom right reproduces the models in the theoretical HR diagram for comparison.}
\label{fig4}
\end{centering}
\end{figure*}

Mass correction factors (solid lines) for solar-type stars are as large as 1.3$\times$, but decrease with inferred age since spotted and un-spotted evolutionary tracks begin to overlap near the horizontal Henyey track (e.g. the 0.9\msun\ tracks). Mass errors increase substantially towards the fully convective regime, where spotted star masses can be under-estimated by up to 2.2$\times$, and never less than 1.7$\times$, regardless of age. The larger errors for low mass stars result from the stronger impact of spots on fully convective luminosities, and the temporal stability of the errors results from the long contraction timescale of low mass stars. Age correction factors (dashed lines) are more severe. For solar-type stars, ages are under-estimated by up to 2.5$\times$, and become more accurate as stars get older. At 0.2\msun, ages will be under-estimated by up to a factor of 10, with the largest errors coming at the earliest epochs. This order-of-magnitude shift results from the large fractional errors incurred when considering values near zero -- for instance, an under-estimate of 1 Myr for a 1.1 Myr old star produces a correction factor of 11. They also relate to the decreasing luminosity of the deuterium birth line with mass: spotted stars are born at a higher luminosity at fixed \teff, and thus lie above un-spotted stars at fixed age (Fig. \ref{fig3}). Consequently, spotted stars interpreted with un-spotted tracks will appear far younger than their true ages, thus requiring a correction factor much larger than unity. We note that the magnitude of the age corrections depends somewhat on how one defines $t = 0$, but that mass correction factors are largely insensitive to this choice.

We note an interesting prediction of these models. In our physical scenario, proto-stars travel up the deuterium birth-line during the assembly phase, slowly gaining mass through accretion. They are prevented from substantial contraction by deuterium burning feedback, and prevented from persisting in a higher luminosity state by the short Kelvin-Helmholtz time-scales \citep[e.g.][]{baraffe09}. The region of the HR diagram above the standard deuterium birth line therefore represents a ``forbidden region'', in which stars with low spot coverage may only reside for a short time. By contrast, heavily spotted stars naturally occur in this location due to their intrinsically lower average surface temperatures. For example, the spotted 3 Myr isochrone in Fig. \ref{fig3} lies entirely above the un-spotted birth line below $\sim 0.5$\msun. Such stars persist in the forbidden region for several tens of Myr, given their slow contraction rates, and should therefore dominate the population coolward of the standard deuterium birth line. This feature provides a direct observational probe of the prevalence of spotted stars in young clusters.

\subsection{Spot effects in the color-magnitude diagram}\label{sec4.2}

\begin{figure*}
\begin{centering}
\includegraphics[width=5.5in]{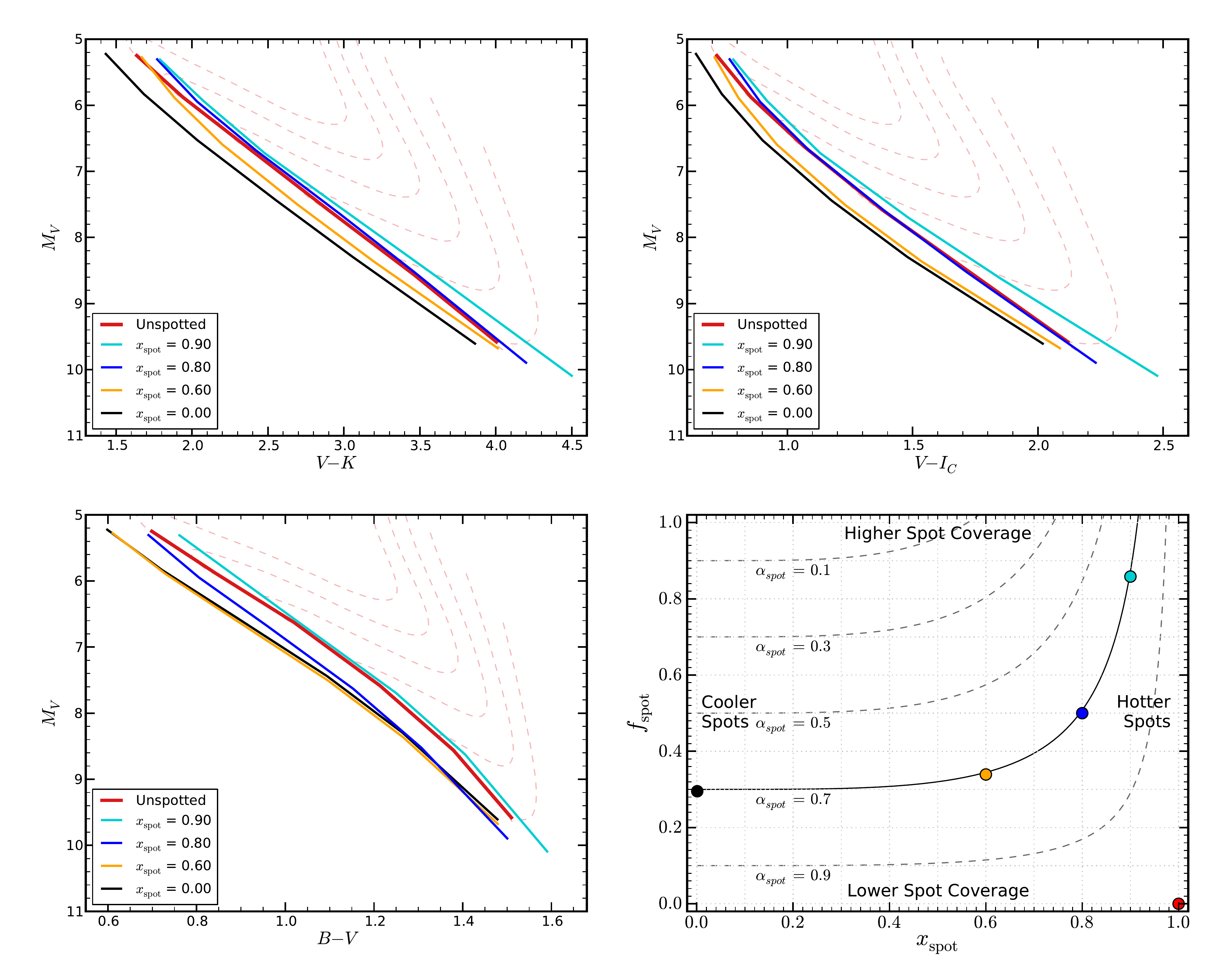}
\caption{The effect of spot filling factor and temperature on the color-magnitude locations of spotted models with the same spot blocking parameter $\alpha_{spot}$. Darker spots cause models to appear bluer, as the hot, un-spotted photosphere progressively dominates the total emission. The bottom right panel shows surfaces of constant flux redistribution in the plane of \fspot\ vs. \xspot, and the colored dots represent the different models shown in the other three panels.}
\label{fig5}
\end{centering}
\end{figure*}

In $\S$\ref{sec4.1}, we drew \teff\ and luminosity directly from our models to predict HR diagram shifts. While valid if \teff\ and L are known, this approach neglects an important fact: spotted stars are inhomogeneously hot across their surface. This will have implications for the spectrum of emission from the photosphere, and may introduce color anomalies relative to pristine stars of equal \teff. To investigate this issue, we transform our models into color space using a two-zone model, where the emission is the area-weighted composite of the SEDs from the surfaces of temperature $T_{amb}$ and $T_{spot}$, whose values are derived from the model \teff\ using Eq. \ref{eqn1}. This transformation is accomplished by using the bolometric corrections of \citet{an07}. Colors are rarely used to interpret pre-MS stars as numerous complicating factors (circum-stellar discs, heavy extinction, etc.) render them un-reliable. We therefore advance our models to 125 Myr for comparison with data in the Pleiades, but note that analogous effects occur on the pre-MS.

The top left of Fig. \ref{fig4} shows the results for $M_V$ vs. $V - K$. The red line represents un-spotted models from 0.5-1.0\msun, and the dark blue line shows our spotted models, transformed using the two-zone approach (\xspot\ = 0.8, \fspot\ = 0.5). The spotted models appear redward of the standard models, though the shift is not as extreme as in the theoretical HR diagram (shown in the bottom right). This is because the un-spotted surface temperature is higher than the model \teff, thus enhancing the short wavelength emission and producing bluer colors. These models are compared to Pleiades data from \citet{kamai14}, supplemented by rotation data from \citet{hartman10}. Pleiads rotating faster than a two day period are shown as open circles, and slower stars are shown as black dots. The slow rotators generally agree with our un-spotted models, but the fastest rotators are typically more red by $\sim$0.1-0.2 magnitudes. As our spotted models also appear more red in this color band, this supports the interpretation that spots are affecting the color of the most active cluster members \citep[e.g.][]{stauffer03,jackson14a}, though we under-predict the magnitude of the shift. The models of \citet{jackson14a} found that spots induce a blue-ward shift in this plane, in contrast to the red-ward shift we find. This discrepancy may be related to differing choices of the spot temperature contrast (see below).

Next, we present isochrones in two additional color coordinates: $M_{V}$ vs. $V-I_C$, and $M_V$ vs. $B-V$. As can be seen, the spotted models appear bluer in shorter wavelength colors. The spotted isochrone lies nearly on top of the un-spotted isochrone in $V-I_C$, and completely blue-ward of it in $B-V$, similar to the behavior of the rapidly rotating data. This is due to the progressively stronger sampling of the Wein tail as one considers shorter wavelengths. It has long been known that the K dwarfs of the Pleiades are blue in $B-V$, and red in $V-K$, compared to isochrones calibrated on older stars. \citet{stauffer03} attributed this behavior to spots and plages brought on by magnetic activity, which respectively enhance the long and short wavelength emission. Our models suggest that qualitatively similar behavior is possible without plages, as the extra blue light can result from a slight increase in the temperature of the un-spotted photosphere. In reality, there are several effects we have neglected in our models which impact the detailed location of CMD isochrones, such as plages, faculae, and other activity-related phenomena. A more complete and detailed model is therefore necessary to make definitive quantitative statements.

These two-zone models were calculated using the same combination of \fspot\ and \xspot. As discussed in $\S$\ref{sec2.2}, these parameters are degenerate from a modeling perspective, since cooler spots with a lower surface area can redistribute the same amount of flux as warmer spots with a higher surface area. This degeneracy is graphically represented in the bottom right of Fig. \ref{fig5}; the curved lines denote surfaces of constant flux redistribution (our base case implies $\alpha_{spot} \sim 0.7$). However this modeling degeneracy does not require that the parameters be observationally degenerate, as the spectrum of photospheric emission depends on the temperatures and relative fractional coverage of the two zones.

\begin{figure*}
\includegraphics[width=7.0in]{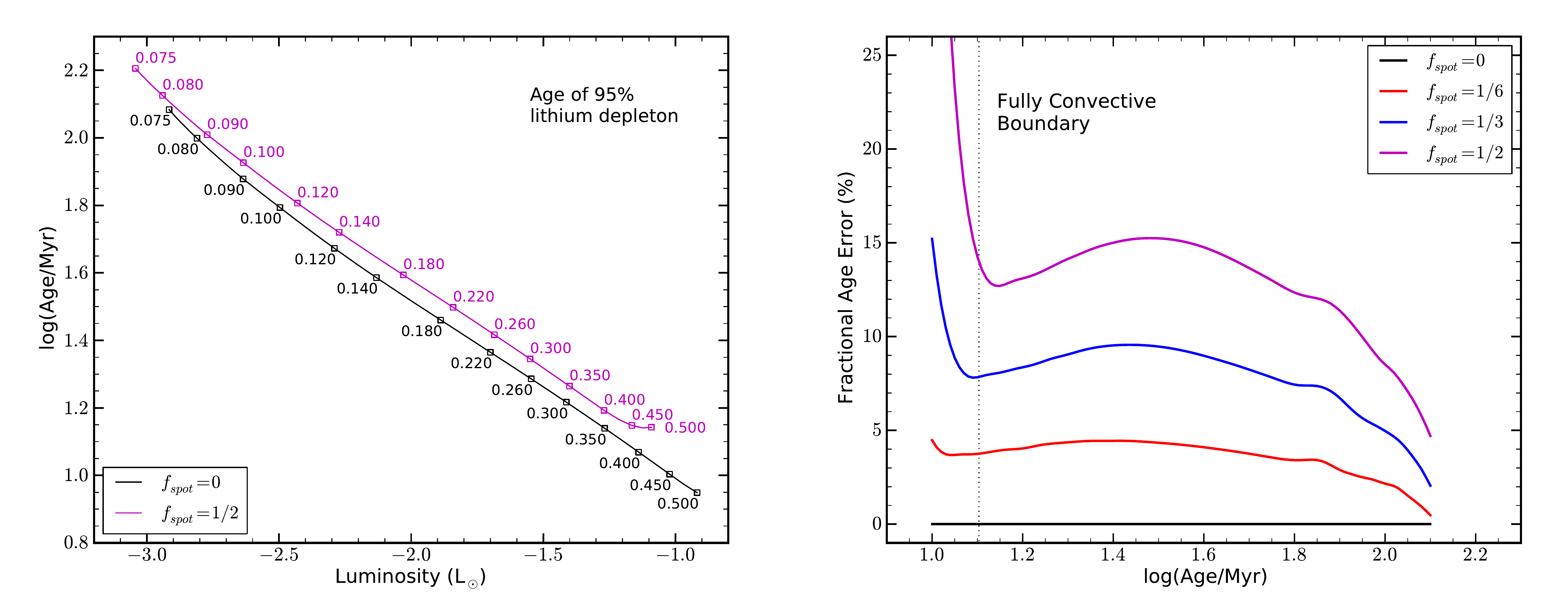}
\caption{The impact of spots on the accuracy of cluster ages obtained through the lithium depletion boundary technique. \textit{Left:} Lines of constant spot filling factor denote the age at which 95\% of the primordial Li is destroyed from models with a given luminosity. Spotted models lie above the un-spotted case, demonstrating that spots delay the onset of Li destruction, thus biasing determinations towards younger ages. \textit{Right:} The fractional error on the age determination as a function of true age, for three different filling factors. The errors are approximately half of the spot blocking parameter $\alpha_{spot}$ (0.3, 0.2, and 0.1 respectively for purple, blue, and red) up to $\sim$60 Myr, and drop off thereafter.}
\label{fig6}
\end{figure*}

We therefore present CMD isochrones calculated with different degenerate combinations of \xspot\ and \fspot\ in Fig. \ref{fig5}. Each panel shows our un-spotted isochrones in red, and isochrones derived through the two-zone approach for four degenerate choices of \xspot\ and \fspot\ (see the colored dots in the bottom right panel). In each case, we see that lower spot temperatures (lower \xspot) give rise to bluer emission, since the decreasing flux of the spots leads to increasing dominance of the global emission by the hot ambient surface. This effect saturates at sufficiently low \xspot\ and asymptotes to totally black spots. The saturation value is a strong function of the color bands: for example, \xspot\ = 0.6 models are nearly indistinguishable from black spots in $B-V$, while they remain separated in $V-I_C$ and $V-K$. In the limiting case of black spots, the entire flux budget originates from the ambient regions, leading to bluer emission in all bands. This behavior is consistent with the findings of \citet{spruit86}, who considered the effect on colors of black spots (see their fig. 2), but did not explore the influence of temperature contrast.

The fact that the observed emission of spotted stars is a composite of two zones may hold crucial implications for the spectra of young stars. Ionization fractions will be non-uniform across the surface, the strength of Zeeman broadening and splitting will differ wildly from location to location, and the cool spotted regions may permit molecular species which are too fragile for the hot photosphere. Given these effects, it is clear that the observed spectrum will be different than a pristine star of equivalent \teff, which may complicate the inference of spectroscopic stellar parameters and metal content. Furthermore, these isochrones show that multi-band photometry holds great value as a tool for breaking the degeneracy between \fspot\ and \xspot, which limits the conclusions of many observational studies of starspots. For example, a given spot modulation signal may increase due to greater (non-axisymmetric) spot coverage, or due to cooler spots. Combining spot modulation with isochrone comparisons may help to constrain the relative contribution of spot area and spot temperature. While we have exhibited the importance of this effect, a full exploration of requires additional important stellar physics as described above, and is therefore outside our scope.

\section{Effect on lithium depletion and cluster dating}\label{sec5}

Lithium has long been used as an indicator of youth in late-type stars. Li is destroyed at $T_{Li} \sim 2.5 \times 10^6$ K for typical stellar densities, and consequently depletes from the surface of stars when the base of the convective envelope reaches this temperature on the pre-MS. For stars more massive than $\sim$0.5\msun, the convection zone retreats and cools below $T_{Li}$ near the end of the pre-MS, thus freezing out the surface Li abundance and imprinting a characteristic mass-dependent pattern at the ZAMS. For stars between 0.5\msun\ and the sub-stellar boundary, Li in the outer layers will be completely destroyed before reaching the MS. Li is also destroyed for some objects below the sub-stellar boundary (0.08\msun$ < M < $0.06\msun), but lower mass brown dwarfs never reach $T_{Li}$ in their interior, and thus do not destroy their Li. As the rate of burning is extremely sensitive to temperature ($\propto T^{20}$), any process that impacts the internal structure of stars may have implications for Li abundance predictions. In this section, we explore the impact of spots on Li destruction in two mass regimes: fully convective stars, which have been widely employed to measure cluster ages, and FGK dwarfs, whose depletion properties have long been at odds with standard theoretical predictions.

Perhaps the most important utility of Li is the measurement of young cluster ages through the lithium depletion boundary (LDB) technique \citep{basri96,bildsten97}. When the central temperature of a fully convective star reaches $T_{Li}$ on the pre-MS, Li is destroyed throughout the star in just a few Myr. The rate of Hayashi contraction, and thus the precise age at which Li destruction occurs, depends strongly on the masses of individual stars. As a consequence, there exists at a given age a critical mass below which stars retain their primordial abundance, and above which Li has been annihilated. Identification of this transition mass, and the subsequent inference of cluster age by comparison with stellar models, comprises the LDB technique. This method has been favored by many authors, due to its relative insensitivity to errors in the physical inputs of stellar evolution codes \citep[$\sim 3-8$\%][]{burke04}. However, \citet{jackson14b} recently showed that because spots reduce the rate of pre-MS contraction (similar to our findings), the age at which $T_{Li}$ is reached in the interior of fully convective stars increases, introducing a bias into LDB ages inferred from spotted stars. In the following, we perform a similar analysis with our evolution code, and compare the results in $\S$\ref{sec6.2}.

To quantify this effect, we present models of $M < 0.5M_{\odot}$ in the left panel of Fig. \ref{fig6}. The lines in this figure represents the age at which 95\% of the primordial Li abundance is destroyed as a function of bolometric luminosity in un-spotted (black line/squares) and spotted (magenta lines/squares) models. The fractional delay in the onset of Li destruction is greater than the fractional luminosity reduction, implying that derived ages will be \textit{under-estimates.} The right panel of Fig. \ref{fig6} shows the fractional age error incurred by using un-spotted evolutionary models to measure the lithium depletion boundary from stars with three different spot filling factors, as a function of the true age. The measurement precision remains roughly constant from $\sim$15 to 60 Myr, and decreases sharply thereafter, due to the progressively stronger $L$ reduction in lower mass stars ($\S$\ref{sec3.1}). For our maximal spot coverage, the fractional age error begins around 13\% at 15 Myr, peaks at $\sim 15$\% at 30 Myr, and decreases below 5\% after 100 Myr.

These numbers indicate errors for a uniformly spotted population. If spot properties vary from star to star, the effect will be to eliminate the one-to-one mapping between mass and Li destruction age, thus blurring out the otherwise sharp depletion boundary. The strength of activity may also depend on spectral type \citep{mohanty03,schmidt15}, particularly near the sub-stellar boundary (SpT $\sim$ M8) in which case different values of \fspot\ should be considered for different masses of the depletion boundary. LDB ages have long been in conflict with ages derived from isochrones, in the sense that LDB ages are systematically higher than MS isochrone fitting, and the spot-induced errors serve to exacerbate this conflict. We note however that the two techniques can be brought into better agreement by the inclusion of convective overshooting on the upper MS \citep[e.g.][]{soderblom14}.

\begin{figure}
\includegraphics[width=3.5in]{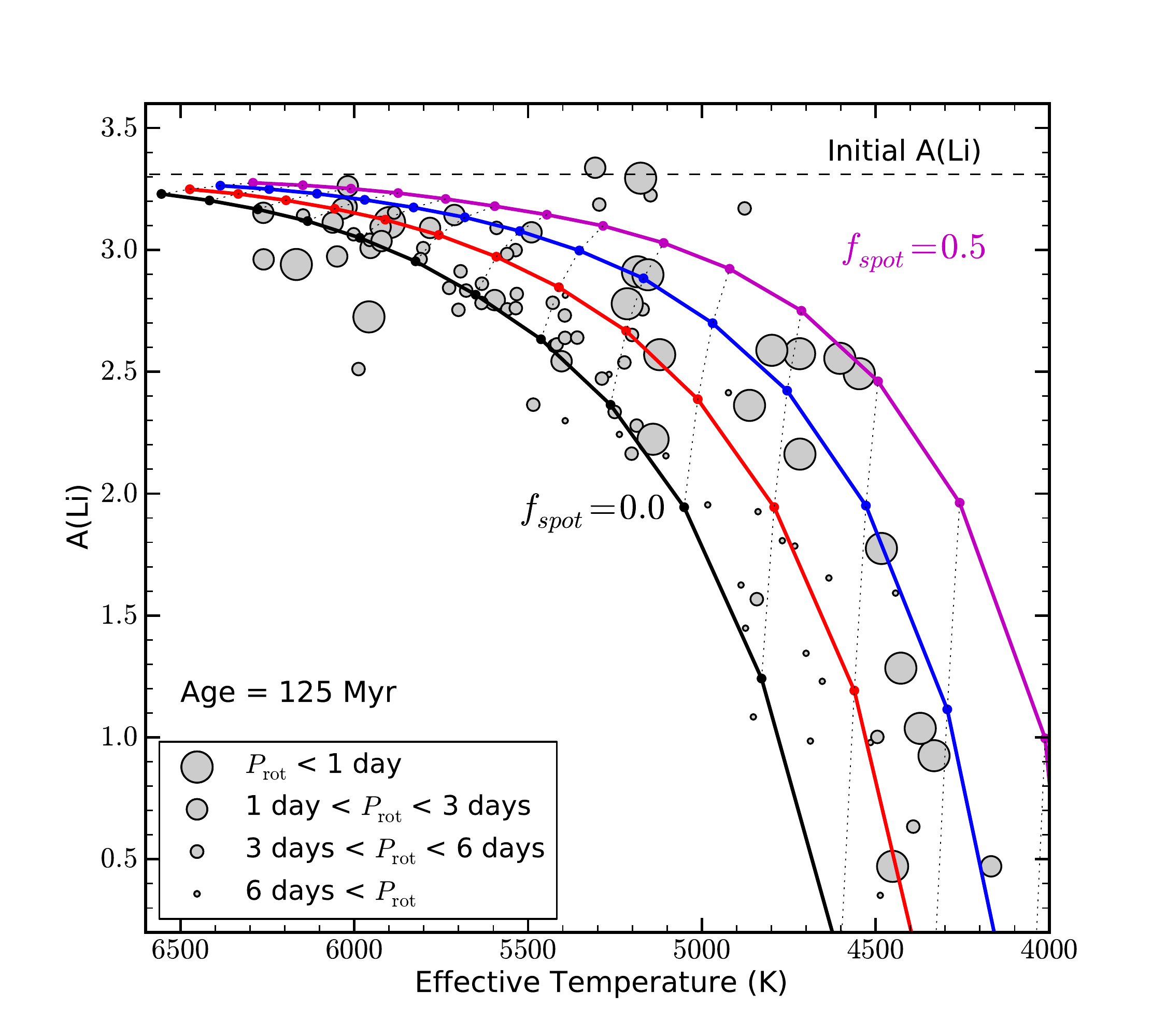}
\caption{The impact of pre-MS spots on Li destruction in partially convective stars. Higher coverage results in a lower pre-MS depletion rate, and thus a higher abundance at the ZAMS. The resulting range and temperature dependence is similar to the Li pattern observed in the Pleiades (grey circles).}
\label{fig7}
\end{figure}

Next, we evaluate the impact of spots on Li destruction in higher mass stars. As discussed above, stars which develop radiative cores cease surface Li depletion when the convective envelope retreats and cools near the end of the pre-MS. This occurs at earlier ages for higher mass stars, so the amount of Li preserved at the ZAMS increases with effective temperature. Standard theoretical models anticipate a strong mass trend with little scatter at the ZAMS \citep{pinsonneault97}. Contrary to this picture, large ($>10\times$) Li dispersions at fixed \teff\ have been identified in the K-dwarfs of several young clusters \citep[e.g.][]{soderblom93,balachandran11}. Such dispersions further correlate with rotation, so that the fastest spinning stars tend to be the most Li-rich. We argued in \citetalias{somers14} and \citetalias{somers15} that if some rotation-correlated mechanism inflated stars relative to standard predictions on the pre-MS, then a rotation-correlated Li dispersion would naturally emerge in open clusters around 10-15 Myr. A dispersion in Li at the ZAMS thus represents a fossil record of non-uniform stellar parameters among pre-MS stars during the epoch of Li burning. We have now implemented a specific mechanism which predicts radius inflation on the pre-MS ($\S$\ref{sec3.1}), and which is known to correlate with rotation in some stars, so we examine its effect on Li depletion in K-dwarfs.

Fig. \ref{fig7} shows the Li abundances of 0.6--1.2\msun\ stellar models at 125 Myr (the age of the Pleiades). Each line represents a different \fspot, given by the key in Fig. \ref{fig6} (\xspot\ = 0.8 in each), and each dotted line connects models of equal mass. As anticipated, different filling factors lead to different Li abundances at the ZAMS, with the dispersion surpassing a factor of ten at some masses. We have also plotted \teff, A(Li), and rotation information for a sample of stars in the Pleiades, assembled in \citetalias{somers15} from data in \citet{soderblom93} and \citet{hartman10}. Point size indicates rotation period, from which the Li-rotation correlation can be readily identified below $\sim 5000$~K. As can be seen, the abundance dispersion in these data roughly corresponds to models in the range \fspot = 0.0--0.45 ($\alpha_{spot} = 1.0-0.73$). This analysis suggests that if rotation correlates with spottiness on the pre-MS as it does on the MS, spot-induced radius inflation naturally produces the Li-rotation correlation in this cluster (see \citetalias{somers15} for a lengthy discussion). This in turn implies an early ($\sim 3 - 15$~Myr) origin for the Pleiades Li dispersion, a hypothesis which may be tested by obtaining Li abundances for clusters of age $\sim$10-15 Myr. Finally, while many stars are more depleted than the un-spotted models above 5000~K, this discrepancy could be explained by rotational mixing \citepalias{somers15}, errors in the input physics of the models \citepalias{somers14}, or could indicate that the Pleiades was born with a sub-solar initial Li abundance ($A($Li$) < 3.31$). 

\section{Discussion}\label{sec6}

We have implemented a simple spot model into our stellar evolution code, and assessed its effect on stars during the pre-MS and on the ZAMS. We have particularly focused on the impact of spots on, 1) stellar structure as a function of mass; 2) the isochrone-derived masses and ages of pre-MS stars; 3) The colors of spotted stars near the ZAMS; 4) the depletion of lithium from stellar envelopes. In this section, we synthesize our results and consider some implications, compare our results with other works who consider magnetic models, and discuss some outstanding issues in both the physical interpretation and modeling of spotted stars.

\subsection{A radius dispersion during the pre-main sequence}\label{sec6.1}

We have shown that if a range of spot properties (i.e. surface magnetic fields) exist at fixed mass and age during the pre-main sequence, the result is non-uniform stellar parameters among the members of young open clusters. But does this scenario accord with observational evidence?

In the case of lithium, we find a conspicuous spread in abundances at fixed temperature among the G- and K-dwarfs of open clusters shortly after the age of pre-MS Li burning. This spread cannot be explained by atmospheric effects \citep[e.g.][]{king10}, by Li depletion in non-magnetic models \citepalias{somers14}, or by the inclusion of rotational mixing \citepalias{somers15}, but naturally emerges from a range of spot properties (i.e. a range of radii) at fixed mass and age during the pre-MS. Furthermore, our predictions are consistent both with the radius data for active eclipsing binaries, and with the anomalous colors of young active stars in the Pleiades, suggesting that our models can reproduce observational results with realistic spot properties, and that magnetic activity continues to inflate rapidly rotating stars on the ZAMS and beyond. Collectively, these results indicate that the differential magnetic properties of stars (correlated with rotation) lead to a dispersion in radius at fixed mass and age, which emerges on the early pre-MS and persists on the MS while stars remain rapidly rotating.

This conclusion is broadly consistent with claims of radius dispersion in pre-MS \citep{cottaar14}, and young MS \citep{jackson09,jackson14a} open clusters, but would appear to be in conflict with the uniform saturation of magnetic proxies among young stars, which could be interpreted to mean that spot properties are identical at fixed mass on the pre-MS. However, we note that saturation is merely an empirical statement about chromospheric and coronal heating, and that a direct correlation with spot properties has not yet been proven (see $\S$\ref{sec6.3.2}). Alternatively, saturation may indicate that other phenomenon, such as mass accretion \citep[e.g.][]{littlefair11}, play a role in generating the pre-MS radius dispersion needed to explain the Li data \citepalias[see the discussion in][]{somers15}, though this scenario still relies on activity to explain the radii of eclipsing binaries and the colors of rapidly rotating Pleiads.

A radius dispersion in pre-MS clusters have several significant implications. First, we have shown that spots make stars appear cooler, and as a result heavily spotted stars coincide on the HR-diagram with younger, less massive stars which are weakly spotted or pristine ($\S$\ref{sec4.1}). The range in masses at fixed HR diagram location can reach a factor of two, while the range in ages can be significantly higher (up to $\sim 3 \times$ above 0.5\msun, and larger for lower masses). Consequently, pre-MS cluster ages inferred by matching isochrones to the locus of HR diagram data will be erroneously low. The magnitude of the age error may match the above-quoted correction factors if stars are uniformly spotted, or may be somewhat less if a range of spot filling factors blurs out an otherwise sharper stellar locus. Furthermore, if stars of equal age have differential spot properties, the inferred ages can differ substantially, leading to the appearance of an intra-cluster age spread. Appreciating the magnitude of this effect is important for inferring the true age spreads within young clusters.

As noted in $\S$\ref{sec5}, the lithium depletion boundary technique also systematically under-estimates the ages of clusters whose members are heavily spotted, by up to 15\% in the most extreme case we considered. As many previous young cluster studies do not consider these magnetic effects, our findings necessitate an upward revision of the age scale for clusters between 1 and 100 Myr. The magnitude of this revision is difficult to anticipate, as the required corrections for each individual study depend strongly on the true age of the cluster and on the mass range probed, as well as the distribution of spot properties among stars in each cluster. A full re-analysis of young cluster CMDs in the context of magnetic models is beyond our current purpose, but would undoubtedly have important implications for, among other things, the duration of circum-stellar disc lifetimes and the accretion timescales of proto-stars, both of which are inferred through cluster ages. 

Second, Fig. \ref{fig3} shows that among spotted stars, the least massive require the largest correction factors. Consequently, in a mono-aged cluster the lowest mass objects will appear to have the youngest isochronal ages, and the hottest stars will appear to have the oldest isochronal ages. Thus in the presence of spots, clusters should show a trend in inferred age with temperature. This precise behavior was noted in the young $\beta$ Pic association by \citet{malo14}: stars with \teff $> 3500$~K appeared to be 15 Myr on average, while cooler stars appear to be 4.5 Myr on average, both of which are younger than the generally accepted age of $\sim 21 \pm 4$~Myr \citep{soderblom14}. They showed that this discrepancy could be ameliorated through the inclusion of strong magnetic fields in the cooler stars, and weaker fields in the hotter stars. Alternatively, consistent spot properties at all masses will naturally impact the age determinations of lower mass stars to a greater degree, and thus can produce just such a disagreement. The relative isochronal ages of cool and warm stars depends on the relative spot properties of the two populations, and therefore provides an empirical constraint on the surface magnetic fields of pre-MS stars.

Finally, the systematic under-estimation of the masses of stars will bias measurements of the initial mass function (IMF) which employ pre-MS clusters. In particular, the average low mass star will require a significant upward revision of its mass, ultimately implying a more top heavy IMF, and a lower number of brown dwarfs. The magnitude of this effect has been discussed in \citet{stassun14}, and we refer the reader there for a more detailed account.

\subsection{Literature comparison}\label{sec6.2}

Heavy spot coverage can enhance stellar radii by up to $\sim$10\% during the pre-MS at fixed age, due in part to the slower contraction and in part to the impact of spots on the pressure conditions at the photosphere. Once on the MS, the peak radius anomalies are mass dependent, with partially-convective stars generally more inflated than fully-convective stars. This pattern is unsurprisingly similar to the findings from \citet{spruit86}, with the exception that our models predict a decrease in $\Delta R$ at fixed spot properties between 0.4-0.8\msun, while \citeauthor{spruit86} predict flat $\Delta R$ in this range. The origin of this discrepancy is unclear, but is probably related to different input physics which cause a substantial shallowing of the mass-radius relationship around 0.7\msun.

Assuming equipartition between the magnetic and gas pressures in the spotted regions (i.e.~$P_{spot,mag} = P_{spot,gas}$) and pressure equilibrium between the spots and the photosphere, our models imply magnetic field strengths of 2.2, 1.6, and 1.1~kG at the ZAMS for 0.2, 0.5, and 1.0\msun, respectively. This is generally consistent with the 1-3~kG range reported for empirical measurements \citep[e.g.][]{cranmer11}. Combined with our spot coverage factor of 0.5, the surface averaged fields \bf\ = 0.6-1.1~kG. This value is similar to the findings of \citet{mullan01}, who require surface strengths of less than 1~kG to inflate fully convective stars by $\sim$3-4\%. They are however somewhat lower than the 2.0-6.0~kG surface fields required by \citet{feiden13,feiden14} to match the observed radii, \teffs, and luminosities of several eclipsing binaries. This quantitative discrepancy is not surprising, as their studies make significantly different assumptions about magnetic field morphology. We note however that the mass dependence they find is qualitatively similar to ours; stronger magnetic fields are required to inflate fully convective stars relative to solar type stars, consistent both with the pattern revealed in Fig. \ref{fig1}, and the \bf\ estimates given above. This suggests that greater inflation at higher mass is a generic prediction of magnetic stellar models.

In $\S$\ref{sec5}, we found that spot-induced errors on cluster ages derived from the LDB technique are large compared to systematic uncertainties in the model inputs \citep[$\sim$15\%, compared to $<$8\%;][]{burke04}. However, the errors we derived are somewhat lower than the $\sim$20\% age errors reported by \citet{jackson14b}, who considered the impact of similar spot coverage on lithium depletion in their polytropic models. These results are in reasonably good agreement, considering the numerous physical inputs required in stellar modeling, but it is interesting to consider the source of this discrepancy. In a polytrope model, the internal gradient $\nabla$ is assumed to be $\nabla_{ad}$ throughout the entirety of convective regions. While this is a good assumption for the vast majority of the interior of fully-convective stars, it is less exact in the thin layer at the surface where the gradient becomes super-adiabatic \citep[e.g.][]{kippenhahn90}. As it happens, starspots chiefly reside near this super-adiabatic layer, and so their effect on the structure of the star will be dependent on the adopted physics of this region. As discussed in $\S$\ref{sec3}, changing the temperature at the boundary of a convective zone leads to a linear shift in $P$ throughout the region. Starspots are placed at this outer boundary in the polytropes of \citet{jackson14b}, so the resulting $L$ shift in the center is approximately equal to the $L$ blocked at the surface. In our standard stellar models, the temperature shift first propagates through the super-adiabatic layer before impacting the convective envelope, leading to a somewhat higher central value of $L$. This in turn impacts the Li depletion age as a function of mass, and thus the inferred ages of clusters. However, we note that the different quantitative results may derive simply from different choices of input physics \citep{burke04}.

\subsection{Observational and physical concerns}\label{sec6.3}

Our approach in this paper has been to adopt a fixed set of spot parameters (i.e. \fspot\ = 0.5, \xspot\ = 0.8 regardless of age), and derive their effect on stellar properties. While such limits are useful, the ideal analysis would tie the temperature, covering fraction, and distribution of spots to the (evolving) structure and properties of their host stars. This physical approach would permit a realistic population model of spotted stars in young clusters, which is desirable for understanding the effect of spots on observables during the pre-MS and MS. Given the present state of the field, such a model awaits superior understanding of the underlying mechanisms governing the behavior of stellar magnetism. We highlight in this section some key outstanding questions whose resolution will improve the capacity of stellar theory to model magnetic fields in young clusters. 

\subsubsection{Spot depth}\label{sec6.3.1}

\begin{figure}
\includegraphics[width=3.5in]{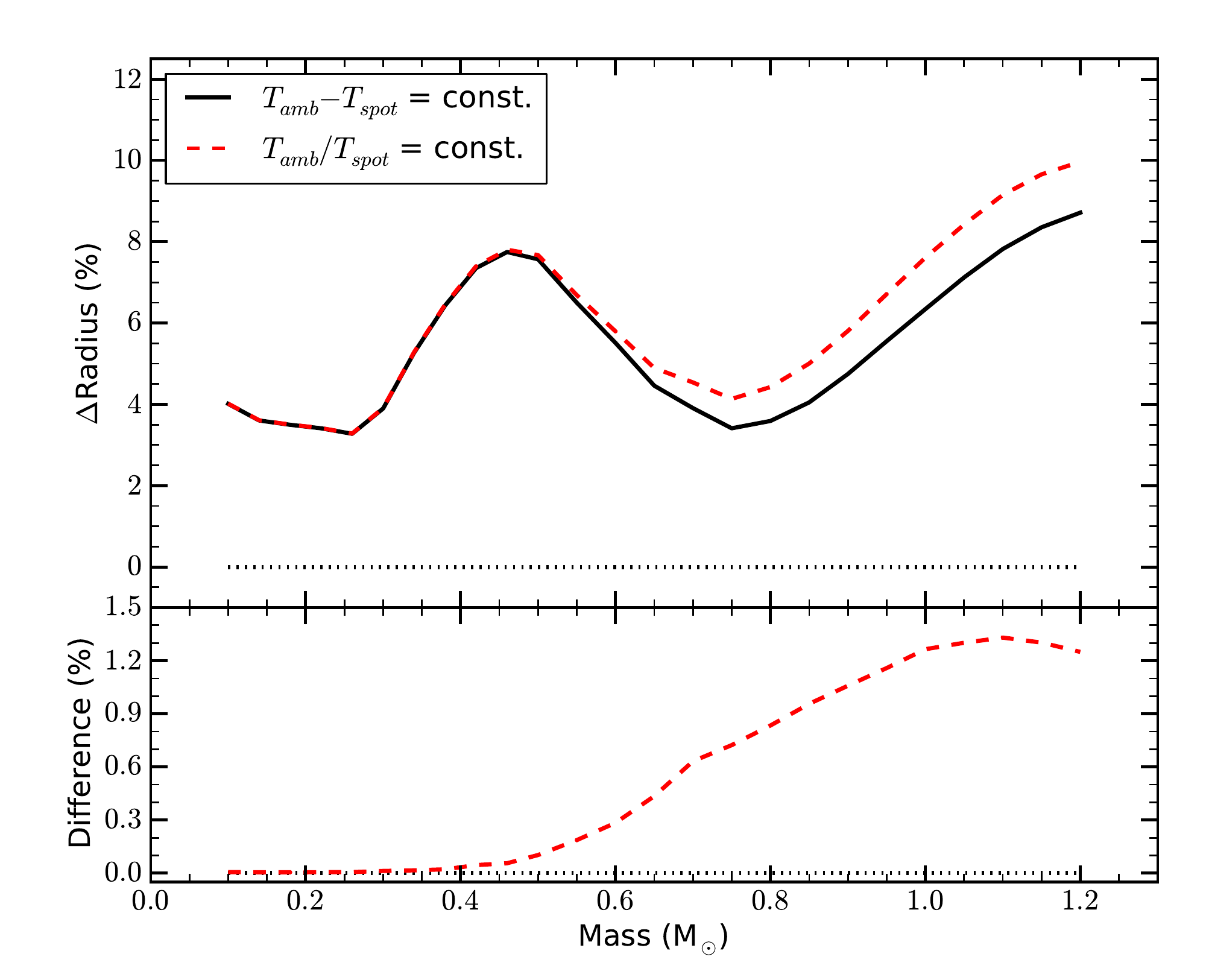}
\caption{The impact of spot depth on radius inflation. The black and red dashed lines represent shallow and deep spot treatments, respectively. The top shows absolute anomalies, and the bottom shows the difference between the anomalies as a function of mass. The difference reaches at most $\sim$1.5\%, and thus impacts detailed calculations, but not the conclusions of this paper.}
\label{fig8}
\end{figure}

Spots extend to a finite depth within the envelopes of stars, but exactly how deep they penetrate, and the details of their subsurface morphology, remain a topic of debate \citep[e.g.][]{rempel11}. In our models, we have assumed that $T_{amb} - T_{spot}$ is constant as a function of depth, a criterion which leads to shallow spots\footnote{As an example, a surface spot contrast of \xspot\ = 0.8 in a 1\msun\ model reduces to less than a 2\% effect at about 3 pressure scale heights below the photosphere, being in the outer $10^{-9}$ and $10^{-3}$ in mass and radius, respectively.}. To assess how this assumption impacts our results, we compare our base case to models which assume a constant temperature contrast with depth ($T_{amb}/T_{spot}$ = constant). This criterion implies very deep spots, since the temperature contrast at the surface is maintained at the base of the surface convection zone.

The comparison appears in Fig. \ref{fig8}. The top panel shows the radius anomalies resulting from the two depth treatments at the ZAMS, and the bottom panel reports the difference between the curves. Deep spots lead to greater radius inflation with a maximum difference of $\sim$1.4\%. This is a small, but non-negligible effect relative to the overall radius anomaly ($\sim$10\%). This suggests that the global impact of spots is chiefly due to their effect on the very outer regions of the star, but is also moderately sensitive to flux redistribution in the envelope. Our constant spot contrast is an extreme limiting case, as it would imply an extremely strong magnetic field (tens to hundreds of MG), and a strong magnetic field gradient, near the base of the convection zone. We therefore expect the true difference to be significantly less than seen in Fig. \ref{fig8}. Despite the marginal differences, shallow and deep spots produce quantitatively similar results, and we therefore conclude that depth effects do not impact our results.

\subsubsection{Magnetic saturation}\label{sec6.3.2}

A key outstanding question in the study of stellar magnetism is the physics underlying magnetic saturation. Numerous studies have determined that once stars are rotating faster than a critical Rossby number \citep[$\sim$0.1; e.g.][]{noyes84,pizzolato03}, classical magnetic field proxies cease to increase with faster rotation. This observation suggests a saturation in the strength of magnetic torques and chromospheric/coronal heating, which are the physical processes that generate the observed proxies. The cause of this saturation remains unclear, but is probably related to a change in the properties of the stellar dynamo when the rotation period becomes substantially shorter than the convective overturn timescale (i.e. low Rossby number).

Observations of young clusters indicate that the vast majority of pre-MS stars are in the saturated regime \citep{preibisch05,argiroffi14}. If all magnetic effects, such as spots, saturate along with the classical proxies, then one would expect uniform spot-properties at fixed mass on the pre-MS, and thus no dispersion in stellar properties due to magnetic activity. However, there is at present no convincing demonstration that spots saturate along with magnetic proxies, and some studies have in fact found evidence that the amplitude of spot modulation varies substantially in the saturated regime \citep[e.g.][]{o'dell95,mcquillan14}, perhaps suggesting disparate spot properties. Furthermore, a uniform spot coverage does not accord with the evidence from Li observations, which indicate a rotation-correlated dispersion in radii at fixed mass and age during the pre-MS ($\S$\ref{sec5}). Uncovering the spot properties of pre-MS stars has been hampered by observational issues, primarily the substantial degeneracies between filling factor, the surface distribution of spots, and their temperature relative to the photosphere \citep{jackson12}. Nevertheless, the connection (or lack thereof) between saturated proxies and spot properties is an essential component of a complete theory of stellar magnetic activity, and must be understood in order to appreciate the impact of spots on the pre-MS.

\subsubsection{Plages}\label{sec6.3.3}

Another confounding effect is the presence of plage regions in stellar photospheres. Plages are the hot analogs of starspots, and radiate light at higher frequencies than the un-spotted surface. This enhanced emission may counteract the total flux suppression caused by spots, and thus reduce the magnitude of the perturbations to stellar parameters that we have outlined in $\S$\ref{sec3}-\ref{sec5}. The size of this correction relies on the difference between the plage enhancement and spot decrement to the total irradiance, which is likely dependent on mass and rotation among other parameters. Furthermore, plage regions provide a source of anomalous blue emission, which has significant effects for the SEDs of active stars \citep[e.g.][]{stauffer03}. These regions are poorly understood at present, so a deeper understanding of their origin, and their correlations with spots on pre-MS and ZAMS stars, is necessary.

\subsubsection{Spot evolution}\label{sec6.3.4}

Finally, in our models we have assumed that spots appear before the deuterium birth line, and persist at the same magnitude for the entirety of the approach to the MS. In reality, spot properties likely change over contraction timescales as the rotation rate and convective properties of stars evolve, as well as on shorter timescales due to the appearance and dissolution of individual spots. A fixed filling factor can approximate the time-averaged effect of spots, but may miss crucial physics occurring on sub-thermal timescales. Furthermore, we have assumed a single temperature contrast between the spotted and ambient regions. In fact, spot temperatures appear to correlate with mass \citep[e.g.][]{berdyugina05}, and may evolve during the pre-MS due to the changing pressure, density, and rotation rate of contracting stars. A superior treatment of these dynamic quantities is necessary for robust spot predictions.

The heliographical distribution of spots may also play a role in determining the efficiency of energy redistribution. For instance, if a large number of small spots are evenly distributed throughout the photosphere, then the entire un-spotted surface will be in close proximity to magnetic regions. By contrast, if the star is populated by a few large spots, then certain ambient regions will be much farther from the closest magnetic region. The average distance of the photosphere from spots could impact the efficiency and uniformity of flux redistribution, and might therefore change the structural effect of starspots. More sophisticated (2-D or 3-D) spotted models may help shed light on the role of spot distribution on surface and sub-surface structure. 

\section{Summary and conclusions}\label{sec7}

Through the inclusion of spots in our stellar models, we have explored the impact of a non-uniform surface on the radii, \teff, luminosity, inferred masses and ages, colors, and Li abundances of pre-MS and ZAMS stars.

We find that a 50\% coverage of 80\% temperature contrast spots (an effective blocking area of 30\% black spots) increases the radii of all stars on the pre-MS by 8-12\%, and by a mass-dependent amount on the MS, which increases from $\sim$3\% in fully-convective stars to $\sim$10\% at 1.2\msun. This behavior is a product of both the impact of spots on the central luminosity of stars, and on the fine structure of the mass-radius relationship. Moreover, these anomalies are similar in magnitude to claimed radius anomalies in detached eclipsing binary systems, suggesting that spots may play a role in the well known radius inflation of young, active low mass stars. 

In our models, these radius anomalies are accompanied by corresponding decrements to the luminosity, which are large in fully-convective stars and small in stars with thin envelopes, and to the \teff, which are weakly sensitive to mass. These perturbations are physical, and displace stars from their fiducial location in the HR diagram. This effect has several implications:

1) Masses derived from spotted stars using pre-MS isochrones will be systematically low by up to $2 \times$ at 3~Myr, and ages will be systematically young by up to $10 \times$ for the lowest mass stars. We quantify the resulting errors as a function of mass and age, and present correction factors for spotted stars.

2) Spotted stars will be preferentially displaced towards the upper right of the standard HR diagram; in particular, they may naturally appear above the un-spotted deuterium birth line. Detecting spot modulation in this region of the HR diagram therefore presents a direct test of the prevalence of spots on pre-MS stars.

3) A range of masses and ages may be present at fixed HR-diagram location, and low mass stars will generally appear younger at fixed age compared to higher mass stars. Interpreters should be cautious about relaying the fundamental parameters of stars without considering the impact of activity.

4) The observed colors of stars will reflect an area-weighted sum of the emission from the hot and cold regions. We find that spotted stars appear bluer than un-spotted stars in short-wavelength bandpasses, and redder in long-wavelength bandpasses, consistent with observations of (spotted) stars in the Pleiades and other young clusters. Furthermore, the extent of the blue-ward shift depends strongly on the average temperature of the spots, which may help to break observational degeneracies between total spot coverage, surface distribution, and temperature.

Finally, we examined the impact of spots on pre-MS lithium depletion in two regimes: fully-convective stars, and in G and K dwarfs. For the former category, the main interest is measuring the ages of young clusters through the lithium depletion boundary technique. We found that an effective blocking area of 30\% can increase the inferred age by $\sim$ 10-15\% relative to the true age. This bias exacerbates the tension between LDB ages and MS turn-off ages in young clusters such as the Pleiades. The errors we derive are less severe than the results of \citet{jackson14b}, with the difference potentially stemming from the treatment of the super-adiabatic region in the two modeling approaches. For G and K dwarfs, we find that spots inhibit the destruction of Li during the pre-MS by altering the temperature at the base of the convection zone. A range of spot filling factors in our pre-MS models leads to varying degrees of Li destruction and thus a dispersion at fixed mass and temperature on the ZAMS, consistent with observational studies of young clusters. This implies that clusters during the epoch of Li destruction ($\sim5-15$~Myr) are differentially affected by magnetism, imprinting a radius dispersion of $\sim 5-10$\% at fixed mass.

\section{Acknowledgments}

We thank John Stauffer, Lynne Hillenbrand, Gregory Feiden, and the anonymous referee for helpful comments on the manuscript. This work was partially supported by NSF grant AST-1411685.


\end{document}